# Electrometry by optical charge conversion of deep defects in 4H-SiC

G. Wolfowicz[*,1,2], S. J. Whiteley[*,1,3] and D. D. Awschalom[†,1,4]

[1]Institute for Molecular Engineering, University of Chicago, Chicago, Illinois 60637, USA
[2]WPI-Advanced Institute for Materials Research (WPI-AIMR), Tohoku University, Sendai 980-8577, Japan
[3]Department of Physics, University of Chicago, Chicago, Illinois 60637, USA
[4]Institute for Molecular Engineering and Materials Science Division, Argonne National Laboratory, Argonne, Illinois 60439, USA

[*]These authors contributed equally to this work.

[†]Corresponding Author: awsch@uchicago.edu

**Optically-active point defects in various host materials, such as diamond and silicon carbide (SiC), have shown significant promise as local sensors of magnetic fields[1,2], electric fields[3–6], strain[7,8] and temperature[9–11]. Current sensing techniques take advantage of the relaxation and coherence times of the spin state within these defects. Here we show that the defect charge state can also be used to sense the environment, in particular high-frequency (MHz-GHz) electric fields, complementing established spin-based techniques. This is enabled by optical charge conversion of the defects between their photoluminescent and dark charge states, with conversion rate dependent on the electric field (energy density). The technique provides an all-optical high frequency electrometer which is tested in 4H-SiC for both ensembles of divacancies and silicon vacancies, from cryogenic to room temperature, and with a measured sensitivity of $41 \pm 8 \; (\mathrm{V/cm})^2/\sqrt{\mathrm{Hz}}$. Finally, due to the piezoelectric character of SiC, we obtain spatial 3D maps of surface acoustic wave modes in a mechanical resonator.**

The detection of electric fields and charge is critical to a wide range of applications including device characterization[5], mapping electrical potential[12] and electrical quantum metrology[13,14]. Recently, electrometry was demonstrated using the spin state of optically-active point defects, specifically nitrogen-vacancy (NV) centers in diamond, enabling quantum-limited sensitivity with nanoscale spatial resolution[3]. Similar experiments were also reproduced in divacancies (VV) in silicon carbide[7]. Nevertheless, electric fields only weakly interact with the spin state of typical qubit defects by altering the zero-field splitting[7] or hyperfine interaction[15,16]. In contrast, an impurity's charge state, though not coherently controllable, is directly sensitive to the electric and charge environment which perturb the electronic wavefunction, and is drastically modified by defect ionization and recombination[17]. The defect charge state provides a naturally occurring analog of quantum point contacts, single electron transistors or other charge-based electrometry devices[13,14].

Optical detection of charge states can be adapted depending on the defect: for the NV center in diamond, a change from the $NV^-$ to the $NV^0$ provides different emission spectra[18], while in VV or silicon vacancies ($V_{si}$) in 4H and 6H-SiC, only one charge state ($VV^0$, $V_{si}^-$) has a known photoluminescence (PL) spectrum[19,20]. Charge conversion between the various charge states can be efficiently realized by optical pumping at specific wavelengths[17,21,22]. Here we show that the optical charge conversion (OCC) rate between the bright and dark charge states of both VV and $V_{si}$ defects is strongly modulated by the presence of an applied radio-frequency (RF) or microwave (MHz to GHz) electric field, and therefore can be detected through changes in PL. The frequency range of this electrometry by optical charge conversion (EOCC) would be extremely challenging using spin sensing due to limitations in Rabi drive rates. We further demonstrate spectroscopic techniques (frequency and phase resolution) using EOCC as well as its application to three-dimensional microelectromechanical system (MEMS) characterization.

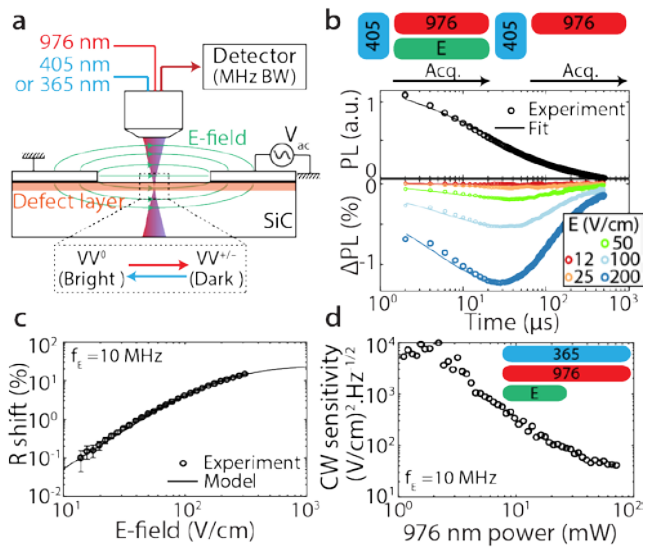

**Figure 1. Electrometry by optical charge conversion (EOCC). a,** Schematic of the optical setup with one reset color (365 nm or 405 nm) and one pump color (976 nm). An RF electric field (r.m.s. amplitude $E$, frequency $f_E$) is applied across a coplanar capacitor with a 17 µm gap. Divacancies are created by carbon implantation immediately below the surface. A fast detector allows for direct detection of a full OCC transient signal in a single measurement. **b,** VV is first reset to its bright state (VV$^0$) by 405 nm illumination, followed by PL detection (top) of the charge conversion toward the VV dark state by 976 nm excitation. Below, the difference between conversion with and without applied electric field (10 MHz) is shown. Data is fit to a stretched exponential function. **c,** Fitted decay rate shifts as a function of electric field. Error bars are 95% confidence intervals. **d,** EOCC sensitivity with continuous (CW) laser (365 nm and 976 nm) pumping as a function of 976 nm laser power. We estimate a 10% error in electric field estimation, corresponding to 20% error in sensitivity (not shown).

In 4H-SiC, OCC of divacancy ensembles requires a near or above bandgap (3.2 eV) excitation to obtain VV$^0$ (bright), while illumination below 1.3 eV pumps the defect toward a dark charge state (VV$^-$ or VV$^+$). We use either 365 nm (continuous) or 405 nm (pulsed) light as reset to VV$^0$ and 976 nm laser as dark state pump[17], with the 976 nm laser also exciting PL from VV$^0$. Fig. 1a presents a schematic of the setup where the two laser beams are focused on divacancies localized near the surface, in between two metal contacts on top of the SiC substrate (see also Supplementary Fig. 1). Applying a voltage across the contacts generates in-plane electric fields orthogonal to the c-axis.

We first characterize OCC transient decays by resetting the charge state with 405 nm illumination followed by a 976 nm pump laser. A fast detector is able to capture a complete transient signal from bright to dark in a single experiment (see Methods section for more details), as shown in Fig. 1b (top). The decay is well fitted by a simple stretched exponential decay $f(t) \propto \exp(-(Rt)^n)$, where $R$ is the characteristic decay time and $n$ the stretch factor, the latter describing the complexity of the charge conversion mechanism ($n = 1$ for simple photoionization and $n < 1$ for competition between ionization, carrier capture and carrier diffusion)[23,24]. We then apply simultaneously an RF electric field along with the 976 nm illumination, resulting in a time-dependent PL variation $\Delta \text{PL}(t, E) = \text{PL}(t, E) - \text{PL}(t, E = 0)$ as plotted in Fig. 1b (bottom), where $E$ is the root mean square amplitude (r.m.s.) of the electric field with corresponding frequency $f_E = 10$ MHz.

While $n$ remains nearly constant (< 1% observed variations), the electric field changes $R$ (see Fig. 1c) according to a phenomenological quadratic dependence with saturation:

$$\Delta R(E) = \Delta R_\infty \langle \frac{(E/E_{\text{sat}})^2}{1 + (E/E_{\text{sat}})^2} \rangle_t \tag{1}$$

where $\langle \rangle_t$ correspond to a time average over an oscillation of the RF electric field, $E_{\text{sat}}$ is the saturation electric field and $\Delta R_\infty$ is the maximum R shift when $E \gg E_{\text{sat}}$. We find for VV in this sample $\Delta R_\infty = 27 \pm 1$ % and $E_{\text{sat}} = 158 \pm 20$ V/cm. It is unclear whether these values are specific to the sample or to the defect itself and additional studies are required. In the first case, EOCC would likely be due to variations in carrier recapture after ionization and would depend on parameters such as mobility or drift velocity. In the second case, $E_{\text{sat}}$ may be directly related to the defect electronic wavefunction and changes in photoionization and capture cross-sections.

Due to the quadratic response given by Eq. 1, EOCC effectively measures the electric field energy density. We define the sensitivity $S$ of this sensing technique for all values of electric fields below saturation as:

$$S = \frac{E^2 \sigma_{\Delta PL}(E) \sqrt{T_{\text{exp}}}}{\Delta PL(E)} \tag{2}$$

where $\Delta PL / \sigma_{\Delta PL}$ is the signal-to noise ratio (using standard deviation) for a given electric field and $T_{\text{exp}}$ is the experiment time (i.e. PL integration time). Fig. 1d shows sensitivity values as a function of 976 nm pump power, optimized by using continuous 976 nm and 365 nm illumination while locking-in on the electric field turned periodically on and off. In this sample, we obtain a sensitivity at 10 MHz as high as $41 \pm 8$ (V/cm)$^2/\sqrt{\text{Hz}}$ for an estimated ensemble of $10^4$ VVs within the confocal spot size. Similar sensitivities have been demonstrated with DC and low frequency (< 1 MHz) spin sensing[3,4], though the quadratic dependence in $E$ makes it hard to properly compare. The sample and VV concentration are similar to that used for typical spin experiments, readily allowing EOCC to be combined with other spin sensing techniques.

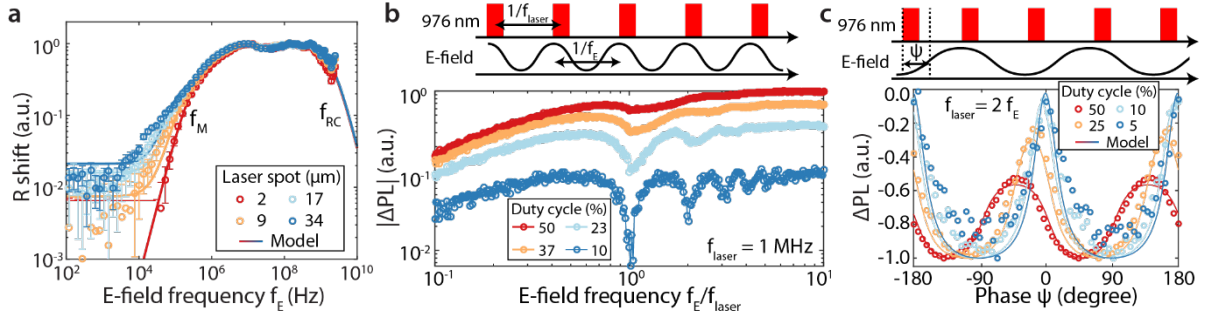

**Figure 2. Frequency and phase domain measurements. a,** Frequency response of the EOCC rate shift (normalized) for different laser spot sizes obtained by changing the microscope focus. At frequencies below $f_M$, the rate shift decays due to a lower effective electric field, while above $f_{RC}$, it decays due to the low-pass characteristics of the experimental setup and device impedance. **b,** Frequency response of the EOCC contrast under pulsed 976 nm light, corresponding to a filter function for RF electric field spectroscopy. The laser pulse periodicity is $1/f_{laser}$. The electric field has a random phase with respect to the laser pulse. The total measurement duration is fixed for all duty cycles. **c,** Phase measurement of the RF electric field using the sequence shown on top. The laser is pulsed at twice the electric field frequency and related by a phase Ψ. Ψ = 0 is defined as the laser pulse coinciding with E = 0; due to finite laser pulse length, this results in asymmetry of the EOCC signal around Ψ = 0. For both **b** and **c**, a 405 nm reset pulse is used before the sequence.

We then characterize in Fig. 2a the frequency response of the EOCC technique by looking at the rate shift ΔR as a function of electric field frequency, from quasi-DC (100 Hz) to 2 GHz. Above 1 GHz, ΔR diminishes as expected from parasitic capacitances of the device (RC filtering), while below 1 MHz, ΔR also decreases possibly owing to the creation of a space charge under illumination and electric field. At low frequency, the optical pumping ionizes the defects, resulting in free carriers that redistribute to compensate the local electric field. At high enough frequency, the carrier distribution never reaches its steady state and the space charge is not created. The characteristic timescale for space charge formation is the Maxwell relaxation time $\frac{1}{f_M} = \frac{\epsilon_0 \epsilon_r \rho}{2}$, where $\epsilon_0$ and $\epsilon_r$ ($\approx 10$ for 4H-SiC) are the vacuum and relative permittivity and ρ is the resistivity[25]. Within this description and using the fitting function for $f_M$ described in the Methods section, we estimate the resistivity at $\approx 10^7$ Ω cm, as expected from typical resistivity values quoted for high purity semi-insulating 4H-SiC wafers. The space charge creation is also expected to depend on the initial charge distribution which we effectively modify by increasing the laser spot size (Fig. 2a). The fit works well in all cases, and in particular for large spot sizes the low frequency rate shift is non-zero (above noise level).

Having characterized the EOCC frequency response, we now demonstrate the ability to resolve the frequency and phase of the applied RF electric field as shown in both Fig. 2b and c., respectively. These experiments are enabled by pulsing the 976 nm pump light with a given frequency $f_{laser}$ and duty cycle (pulse duration). First, we fix $f_{laser}$ while sweeping $f_E$ with a random initial phase between the two frequencies. This sequence measures the effective filter function of the pulse sequence, and shows dips of decreasing intensities for $f_E$ equal to increasingly higher harmonics of $f_{laser}$ (Fig. 2b). The dips arise from the light pulse always overlapping with equal electric field values when $f_E$ matches a harmonic of $f_{laser}$; the RF electric field effectively becomes static in this condition and the EOCC signal diminishes as expected from Fig. 2a. The effect is gradually more prominent for decreasing duty cycle as the filter function sharpens. For phase resolution (Fig. 2c), we fix the relative phase Ψ between the laser pulse and electric field oscillations and set $f_{laser} = 2f_E$. Alternating light pulses encounter electric fields with alternating signs but equal amplitude depending on the phase Ψ, and sweeping Ψ hence maps the time evolution of the electric field ($E^2$) oscillation.

The model in the figure is calculated without any free parameters using Eq. 1 and the overlap between the electric field wave and the laser pulse.

The electrometry technique we have outlined is broadly applicable to other defects with known charge dynamics. For example, $V_{si}$ can be optically charge converted and, unlike VV, is optically active even at room temperature. The combination of 365 nm (pumping to a dark state) and 785 nm (pumping to a bright state) lasers allows for OCC[17] and therefore the application of EOCC as shown in Fig. 3a,b. Under continuous illumination at both wavelengths, the electric field modifies the $V_{si}$ PL for temperatures ranging from 5 K to 350 K. The EOCC contrast is present at all temperatures, though it is strongly reduced above 30-77 K; this behavior could be explained by the thermal activation of shallow impurities or capture barriers. We do not compare here $V_{si}$ and VV sensitivities as the experimental setup and the sample were only optimized for VV defects.

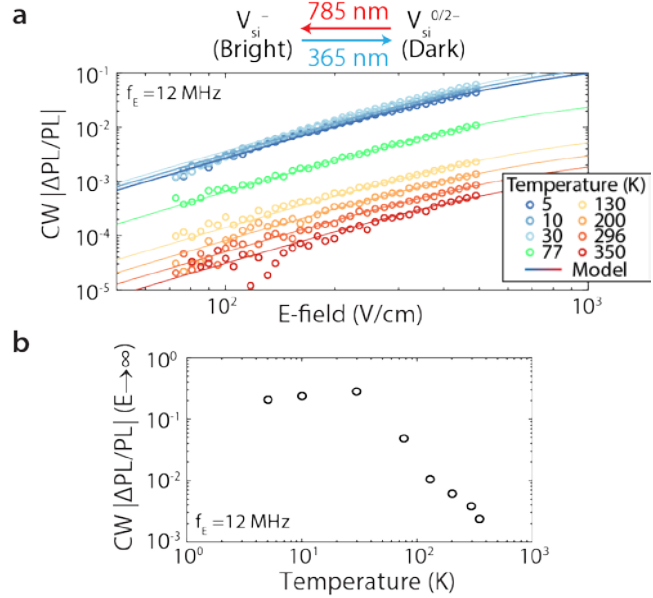

**Figure 3. EOCC in silicon vacancies ($V_{si}$) in 4H-SiC. a,** EOCC contrast $|\Delta PL/PL|$ as a function of electric field for temperatures ranging from 5 to 350 K. Lines are fit to Eq. 1, with $E_{sat}$ fitted to be $610 \pm 80\ V/cm$ independent of temperature. **b,** Extrapolated EOCC contrast for $E \to \infty$ as a function of temperature, with 95% confidence intervals (not shown) equal to about 5% of the contrast.

To conclude this work, we demonstrate the application of EOCC to map surface acoustic wave (SAW) modes in an electro-mechanical resonator in 4H-SiC. As SiC is slightly piezoelectric, any strain or shear wave simultaneously produces a corresponding measurable electric field. The SAW resonator is displayed in Fig. 4a (see Methods section and Supplementary Fig. 2,3 for further details) with an interdigital transducer (IDT) fabricated on top of a 500 nm AlN layer on top of the SiC substrate. The resonator is composed of Bragg gratings made from grooves in the AlN that act as reflective mirrors, while the IDT couples the electrical drive to the SAW mode. All PL measurements are realized away from the IDT to avoid contribution from the drive electric field. Fig. 3b-d present a longitudinal (x-z) cross-section in the center of a device where there is a window in the IDT, a cut (x) across the AlN grooves and a transverse cut (y) in the central window, respectively. In the cross-section of the window, we observe wave crests separated by half of the SAW wavelength λ (λ = 16 μm, cavity frequency is 421 MHz), as expected from a quadratic response in electric field.

Numerical simulations confirm the contrast to likely originate from the $E_x$ electric field component of the resonator mode (see Supplementary Fig. 4). In panel c, the cut through the grooves show oscillations from the SAW modulated by an exponential decay. The characteristic decay length is measured to be $L = 0.78 \pm 0.03$ µm, and directly related to the reflectivity per grating strip $|r_s| = \frac{\lambda}{4L} = 0.51 \pm 0.02$ %.

A transverse sweep across the central window measured as a function of drive frequency allows for direct observation of various transverse modes of the SAW resonator. Fig. 4d shows modes with 1 to 5 peaks (i.e. electric field extrema); their respective signals are separately integrated, plotted (bottom) and compared with a direct reflection ($S_{11}$) measurement of the cavity. The $S_{11}$ signal only provides the total contribution from all modes, whereas the EOCC technique fully separates each mode in spatial and in frequency domains. Overall, EOCC offers complementary information to common MEMS characterization methods such as laser Doppler vibrometry[26] and various surface techniques (scanning electron microscope, atomic force microscopy, etc.)[12]. In particular, 3D spatial and high frequency (GHz) sensing available with EOCC are much harder to achieve with these alternative techniques.

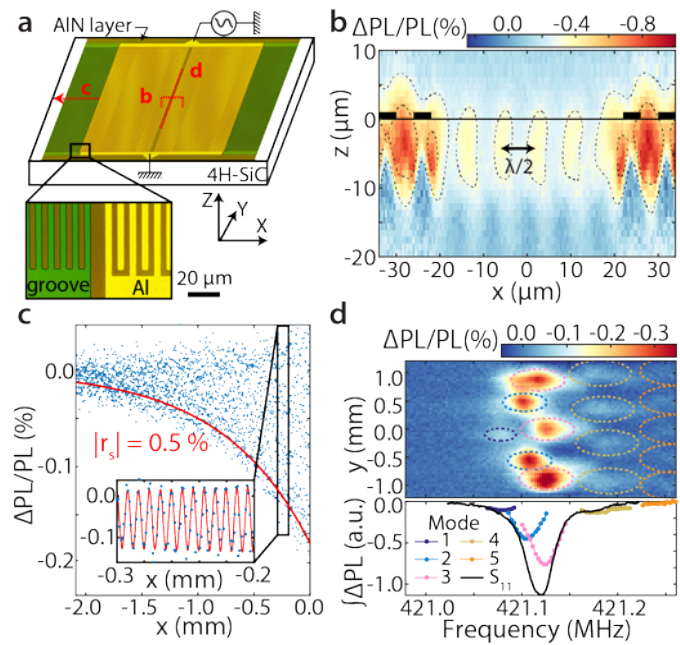

**Figure 4. Surface acoustic wave (SAW) mapping. a,** Schematic of the SAW device with partial microscope images. The red lines are scans with corresponding panel letters. The AlN grooves extend for 5.4 mm on each side. **b,** EOCC contrast of a cross-section (x-z) centered at the window of the resonator. The black line indicates the surface and the black boxes the metal from the interdigitated fingers (not to scale). **c,** EOCC contrast near the AlN grooves (reflectors) allows for direct measurement of their reflectivity $r_s$. $x = 0$ corresponds here to the position of the first groove. In inset, a zoom is shown to emphasis the oscillations from the SAW. **d,** SAW drive frequency as a function of transverse (y) position. 5 modes are observed and integrated according to the colored ellipses. The integrated signals for each mode are shown as a function of frequency and compared to the $S_{11}$ intensity from a Schottky diode. All maps were realized at 0 dBm input power.

In summary, we present a new electric field sensing technique for optically-active defects in 4H-SiC. This method is purely optical and has a quadratic dependence in the applied electric field, i.e. measures the electric field energy density, whose frequency can be as high as a few GHz, likely limited by the experimental setup. Electrometry at such frequencies would be hard to achieve with spin sensing techniques. We further demonstrate spectroscopy in both the frequency and phase domain. The ability to measure electric field vectors (3-axis) could be realized by taking advantage of the non-linear (quadratic) response in electric field. Finally, we demonstrate mapping of a SAW resonator due to the piezoelectricity of SiC, offering a new characterization tool for related MEMS. Further improvements in the sensitivity could be achieved by higher defect densities, which should not drastically affect the charge state dynamics contrary to the spin coherence, or toward high spatial resolution by using single impurities. This technique is likely applicable to defects in other materials, in particular large bandgap crystals such as diamond and other substrates for high power electronics and high frequency MEMS.

**Methods**

**Samples.** The coplanar capacitor device was fabricated on a semi-insulating 4H-SiC commercial wafer from Norstel AB. VV and $V_{si}$ defects were created by carbon ($^{12}$C) implant with a $1 \times 10^{12}$ cm$^{-2}$ dose at 170 keV with a 7° tilt (~300 nm depth), followed by annealing at 900°C in Ar for 2 hours. 10/90 nm of Ti/Au was used for the metal gates. The device design is shown in Supplementary Fig. 1 and has multiple capacitors in parallel, though the laser spot in all experiments was confined to a single capacitor (with 17.1 μm spacing).

The SAW resonator (see Supplementary Fig. 2,3) was fabricated on a semi-insulating 4H-SiC commercial wafer from Cree Inc.. Defects were created by carbon ($^{12}$C) implant with a $1 \times 10^{12}$ cm$^{-2}$ dose at 170 keV with a 7° tilt (~300 nm depth), followed by annealing at 900°C in N$_2$ for 2 hours. 500 nm of AlN was sputtered on the Si face of the wafer by OEM Group Inc. The AlN layer has -40 MPa film stress with a rocking curve for AlN (0002) of 1.52° full width at half maximum (XRD). 150 nm of Al was used for the interdigital contacts (80 finger pairs, with a window in the center equal to 3λ of missing fingers). Al and AlN were etched by inductively coupled plasma (ICP) with 10 sccm Ar, 30 sccm Cl$_2$, 30 sccm BCl$_3$, 50 W bias and 400 W ICP power. The grooves in the AlN were patterned by optical lithography and etched 270 nm deep.

**Experimental setup.** Samples are mounted on printed circuit board inside a closed-cycle cryostat. All measurements were realized using a single confocal microscopy setup (50x objective) with optics optimized for near-infrared. For VV, OCC was realized using a 365/405 nm and 976 nm laser diode, with 976 nm simultaneously exciting the VV$^0$ photoluminescence (> 1000 nm). For $V_{si}$, OCC was realized using the same 365 nm and a 785 nm laser diode, with 785 nm simultaneously exciting the $V_{si}^-$ photoluminescence (875-1075 nm filtering). For pulsed laser experiments, the 976 nm laser was modulated using an acousto-optic modulator (<< 100 ns rise time) while 405 nm was directly modulated by a current driver (250 kHz).

Detection was realized using two separate configurations. For continuous measurements, an InGaAs photodiode with 1 kHz bandwidth was combined with a lock-in amplifier set at the frequency of the electric field amplitude modulation or switching (typically 400 Hz). For direct transient detection, an InGaAs avalanche photodiode (Thorlabs APD410C) in linear regime (M factor = 20) with 10 MHz bandwidth was used with a fast (125 MHz) acquisition card. Transient signals were acquired with a 50 MHz sampling rate and binned into 2 μs samples; differential measurements for ΔPL are numerically processed during acquisition.

All maps were taken using a 3-axis linear stage. AC electric fields are generated by two separate sources below and above 40 MHz. Above 40 MHz, the input power is calibrated to be flat across all frequencies by measuring the reflected power from the sample with a Schottky diode.

**Low frequency response.** In order to understand the low frequency response, we tested other device geometries such as a simpler coplanar capacitor design and a coplanar waveguide, as well as different 4H-SiC wafers. The effect of device impedance at low frequencies is disregarded as we did not observe any change using these various device configurations.

The low frequency behavior can be modeled by an effective electric field with the following linear frequency response:

$$E(f_E) = E_{\text{HF}} \left| 1 - c \times \frac{1 - if_E/f_{\text{M}}}{(1 + a - if_E/f_{\text{M}})(1 - b - if_E/f_{\text{M}})} \right| \qquad (3)$$

where $E_{\text{HF}}$ is the electric field value at high frequency, $i$ is the imaginary unit and $a$, $b$, $c$ and $f_{\text{M}}$ are free parameters. Though this equation is purely phenomenological, it is similar to theoretical calculations for the conductivity response from the creation of a space charge (but for different conditions from our experiments)[25]. In this case, the parameter $f_{\text{M}}$ is the Maxwell relaxation rate and $a$ and $b$ are related to the inhomogeneous distribution of free carriers due to photoionization.

**Acknowledgments**

We thank Christopher Anderson, Joseph Heremans, Alexandre Bourassa, Brian Zhou and Berk Diler for insights, discussions and reviewing the manuscript. This work was supported by the ARL OSD QSEP program and the NSF EFRI 1641099 grant. This work made us of the MRSEC Shared User Facilities (NSF DMR-1420709) and the Pritzker Nanofabrication Facility (SHyNE Resource, NSF ECCS-1542205) at the University of Chicago. G.W. acknowledges support from the University of Chicago/Advanced Institute for Materials Research (AIMR) Joint Research Center. D.D.A. is supported by the US Department of Energy, Office of Science, Basic Energy Sciences, Materials Science and Engineering Division at Argonne National Laboratory.

**Author contributions**

G.W. and S.J.W. performed the experiments. S.J.W. designed and fabricated the samples. D.D.A. advised and coordinated all efforts. All the authors contributed to analysis of the data, discussions and the production of the manuscript.

**Additional information**

Competing interests: The authors declare no competing financial interests.